\documentclass[%
aps,
prb,
reprint,
superscriptaddress,
]{revtex4-2}

\usepackage{xcolor}
\usepackage{textcomp}
\usepackage{graphicx} 
\usepackage{float}
\usepackage{epstopdf}
\usepackage{txfonts}

\usepackage[colorlinks,citecolor=blue,linkcolor=blue]{hyperref}

\newcommand{\CNN}{Centre de Nanosciences et de Nanotechnologies, CNRS, Universit\'e Paris-Saclay, 91120, Palaiseau, France}
\newcommand{\LabSTICC}{LabSTICC, CNRS, Universit\'e de Bretagne Occidentale, Brest, France}

\begin{document}

\title{Spin-wave softening across the uniform-to-stripe domain transition in iron garnet film}

\author{Titiksha Srivastava}
\email{titiksha.srivastava@c2n.upsaclay.fr}
\affiliation{\CNN}
\author{Victor Leroy}
\affiliation{\CNN}
\author{Nessrine Benaziz}
\affiliation{\CNN}
\author{Nathaniel Findling}
\affiliation{\CNN}
\author{Ludovic Largeau}
\affiliation{\CNN}
\author{Jamal Ben Youssef}
\affiliation{\LabSTICC}
\author{Jean-Paul Adam}
\affiliation{\CNN}
\author{Thibaut Devolder}
\affiliation{\CNN}
\author{Joo-Von Kim}
\affiliation{\CNN}

\date{\today}

\begin{abstract}
Spin-wave spectra across transitions between uniform and textured phases can offer deep insight into both symmetry-breaking physics and self-assembled magnonic bands. However, experiments require a material platform that combines low damping, well defined textures, and spectroscopic access. Here, we study Bi-doped iron garnet film with perpendicular magnetic anisotropy (PMA), which undergoes a uniform to stripe domain transition as a function of in-plane field. While real space imaging using magnetic force microscopy shows field-reorientable stripe domains formed along the in-plane field direction, reciprocal space study using thermal microfocused Brillouin light scattering ($\mu$-BLS), reveals the softening of a low-frequency spin-wave branch near the transition and the appearance of additional modes in the stripe-domain state. Calculated dispersion relations identify a finite-$k$ softening in the Damon--Eshbach geometry ($\mathbf{k}\perp\mathbf{M}$), matching the stripe periodicity at the transition. In addition, a $\mu$-BLS spectral model reproduces the measured  mode frequencies and relative intensities at selected fixed fields. Micromagnetic simulations capture the field-driven formation of the stripe state and reproduce the experimental thermal $\mu$-BLS spectra. Our findings establish BiYIG with PMA as a model low-damping platform for studying spin-wave freezing, stripe-domain modes, and reconfigurable magnonic band structures.

\end{abstract}

\maketitle

\section{Introduction}

Spin-wave modes are collective excitations in magnetic materials that are intrinsically tied to the underlying magnetic ground state. In extended thin films with uniform magnetization, spin waves are extended plane waves with continuous dispersion. These include the uniform precession mode described by Kittel, as well as propagating modes such as Damon--Eshbach (DE) and backward-volume (BV) spin waves, which differ in the relative orientation of the wavevector and equilibrium magnetization~\cite{kalinikos86}. In contrast, in nano-patterned samples or magnonic crystals, engineered geometries impose artificial periodic potentials that modulate spin-wave propagation, resulting in band folding, energy gaps, and localized eigenmodes, thus allowing spin-wave control for devices~\cite{kruglyak10,krawczyk14}.

An exciting alternative to nano-patterned samples is provided by magnetic textures. These textures emerge spontaneously in certain magnetic systems as stable or metastable states, depending on the balance between exchange, perpendicular magnetic anisotropy (PMA), dipolar and Zeeman energies. Examples include skyrmion lattices~\cite{fert17}, vortices, magnetic bubbles, and stripe domains~\cite{IVANOV89, Garnier20}, where the periodicity is self-organized. Recently, these textures have taken centre stage in magnonics, as they offer rich magnon spectra, including localized and collective modes, as well as possibilities for spin-wave control without the need for complicated lithography~\cite{yu21, argyle83, ebels01, gubbiotti12, garst17, mochizuki12, montoya17, mruczkiewicz16, onose12, lonsky20, satywali21, schwarze15, Srivastava23}.

Among such systems, stripe domains constitute a particularly simple and versatile example of a self-organized periodic magnetic medium. They consist of alternating domains with opposite out-of-plane magnetization components, separated by domain walls, and their period and orientation can be tuned by film thickness, strain, magnetic field, and magnetic history~\cite{HubertSchafer98}. They have been shown to exhibit a wide spectrum of localized and delocalized modes, spanning gigahertz to sub-gigahertz frequencies~\cite{prestwood25, camara17, Dhiman24, vukadinovic01, banerjee17, Ebels02, Ramesh88, Franck03}. At the same time, stripe-domain systems can provide a model platform for exploring dynamical fingerprints of magnetic phase transitions. In sufficiently homogeneous films the saturated state can become unstable to a finite-wave-vector magnetization modulation, allowing the stripe phase to emerge through the softening of a spin-wave mode~\cite{Leaf06,Jan23}. The amplitude of this modulation then acts as an order parameter, while the stripe-domain period sets the characteristic length scale of the texture. Below the transition, the spectrum reorganizes around the new periodic ground state, providing a route to symmetry-derived excitations such as Goldstone-like and Higgs-like modes.

Recent experiments have provided important evidence for symmetry-derived modes in stripe-domain systems based on metallic films~\cite{grassi22}. However, the reciprocal-space evolution of the spectrum across the transition, namely how spin-wave modes soften, reorganize, and acquire band character as the system evolves from a uniform to a periodically modulated magnetic state, remains difficult to observe due to high damping parameters and material inhomogeneities. Low-damping magnetic insulators with PMA, which can host well-defined and tunable stripe domains, offer a complementary platform to address this question.

In this work, we investigate the spin-wave spectral evolution across the  saturation-to-stripe-domain transition in a 120-nm-thick BiYIG film with PMA. By combining magnetic force microscopy, thermal microfocused 
Brillouin light scattering ($\mu$BLS), dispersion calculations, and micromagnetic  simulations, we follow both the field-driven formation of a stripe-domain  lattice and the reorganization of the spin-wave spectrum across the transition. The calculated dispersions identify the relevant finite-$k$ softening in the Damon--Eshbach  geometry ($\mathbf{k}\perp\mathbf{M}$), while a thermal $\mu$-BLS spectral model~\cite{Benaziz2025} and micromagnetic simulations reproduce the measured mode frequencies and spectral intensities. Together, these results provide a consistent picture of how the spin-wave spectrum softens and reorganizes as the film evolves from the uniformly magnetized state to the stripe-domain state.

\section{Results}
\subsection{Material growth and structure characterization}
\begin{figure}
	\centering
        \includegraphics[width=\columnwidth]{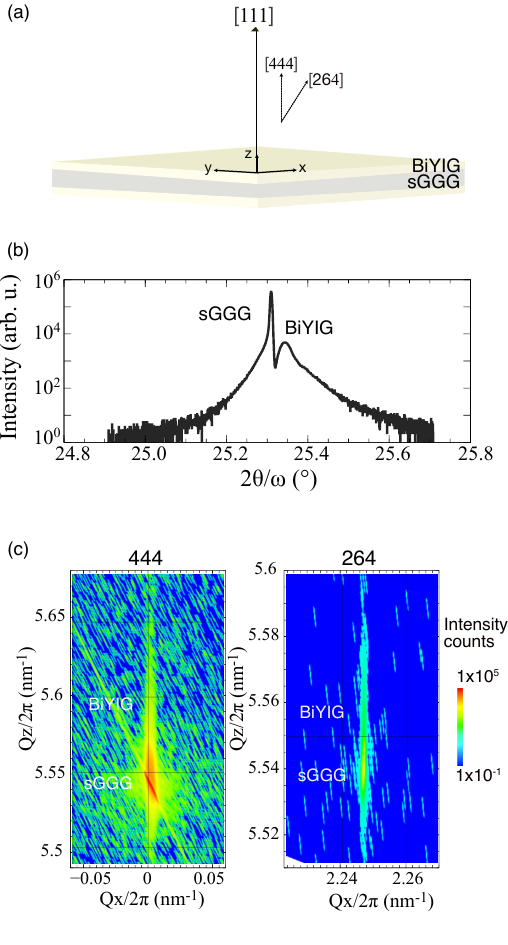}
	\caption{(a) Schematics of Bismuth-doped Yttrium Iron Garnet sample grown on (111)-oriented sGGG substrate. (b) $2\theta-\omega$ X-Ray diffraction scan for BiYIG film. The film and the substrate diffraction peak indicates that the film in in tensile strain (c) Reciprocal space mapping of BiYIG film on sGGG : left symettric (444) and right (264) antisymmetric reflections evidencing pseudomorphic growth. The film peak lies directly above the substrate peak along Qx, confirming full lateral coherence, while its Qz position above the substrate indicates an out-of-plane mismatch of the lattice parameter leading to tensile strain in this case.}
	\label{fig:Structure}
\end{figure}

The BiYIG (Bi$_x$Y$_{3-x}$Fe$_5$O$_{12}$) $120$~nm thick sample was grown by liquid phase epitaxy (LPE) on a (111)-oriented sGGG substrate (shown in Fig~\ref{fig:Structure}a) where $x\sim1.4$.~ Fig~\ref{fig:Structure}b shows high resolution X-ray diffraction (HRXRD) $2\theta-\omega$ scan of the $444$ reflection of the film measured using a PANalytical X’Pert diffractometer equipped with a Cu X-ray tube, parallel beam optics, and a Ge$220 \times 4$ incident-beam monochromator. The diffraction curve exhibits a substrate peak and a relatively well-defined peak for the layer. The relative shift of the BiYIG peak with respect to the substrate peak indicates that the film is under tensile strain. 

To further explore the strain state, reciprocal space mapping (RSM) was performed using a Rigaku Smartlab diffractometer equipped with a Cu rotating anode, a Ge220x2 monochromator on the incident beam and a 2D detector for fast reciprocal space mapping (RSM) acquisition. Two types of RSM are collected around the symmetric 444 reflection, which probes the out-of-plane lattice parameter, and around the asymmetric 264 reflection, which provides information on both in-plane and out-of-plane lattice parameters Fig~\ref{fig:Structure}c. In the RSM, the BiYIG reflection is vertically aligned with the sGGG reflection along the in-plane reciprocal-space direction, indicating that the film is pseudomorphic and fully strained, with no in-plane relaxation. The magnitude of the epitaxial strain is reflected by the separation between the substrate and the film peaks in the reciprocal space.

\subsection{Magnetization statics}
\begin{figure}
	\centering
        \includegraphics[width=\columnwidth]{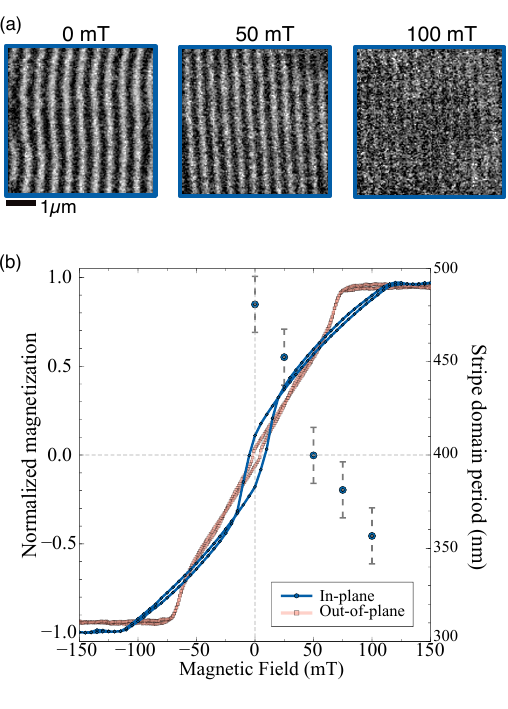}
	\caption{(a) MFM images of BiYIG film at remanence and applied in-plane magnetic fields.  (b) SQUID-VSM magnetization curves for in-plane and out-of plane applied magnetic field and the stripe domain period evolution.}
	\label{fig:StaticMagnetic}
\end{figure}

The magnetic hysteresis loops of the film were characterized by Superconducting Quantum Interference device (SQUID) vibrating sample magnetometry (VSM). Measurements with field applied in-plane (IP) and out-of-plane (OP) are shown in Fig~\ref{fig:StaticMagnetic}(b). The film has a small effective out-of-plane anisotropy as seen from the hysteresis loops. High field-FMR measurements are used to extract the damping parameter $\alpha = 1.2 \times 10^4$ with inhomogeneous broadening less than $0.1$~mT (see Supplementary Material~I). The saturation magnetization was estimated from VSM measurements: $M_s = 160 \pm 10$~kA/m. The effective anisotropy and the uniaxial anisotropy constant is extracted using $\mu_0 M_{\mathrm{eff}}=$$\mu_0 (M_s-H_k) = -7.2\pm 0.5$~mT  where $\mu_0 H_k = 2 K_u/M_s$, hence $K_u=120 \pm 0.11$~kJ/m$^3$. The gyromagnetic ratio ($\gamma/2\pi = 28\times 10^9$~Hz/T). The above values of $K_u$ and $M_s$ lead to a quality factor, $Q \sim 1 $ ($Q = 2K_u/\mu_0 M_s^2$). 

Note that the origin of PMA in BiYIG is attributed to both magneto-elastic strain and growth-induced anisotropy  \cite{soumah18,Fratello86,Gouere2022}. In other words, since the magnetostrictive parameter of BiYIG on sGGG (111) is negative, the tensile strain owing to the lattice parameter mismatch of BiYIG and sGGG substrate (sGGG has a larger lattice parameter than BiYIG), favors an out-of-plane anisotropy. The growth induced contribution, coming from the preferential distribution of Bi${^3+}$ and Y${^3+}$ at non-equivalent dodecahedral sites of the unit cell, adds up to magneto-elastic strain to compensate the dipolar energy thus favouring an out-of-plane uniaxial magnetic anisotropy~\cite{Jamal89,Fakhrul23,Das23}. 

As $Q \sim 1$ we expect a dense stripe structure in our sample with Bloch like domain walls and possible Néel caps for flux closure~\cite{HubertSchafer98}. Fig~\ref{fig:StaticMagnetic}(a) shows stripe domains measured by Magnetic Force Microscopy (MFM) after saturation of the sample by an in-plane magnetic field. The stripes are oriented along the initial applied in-plane saturation field. It is interesting to note that the stripe domain structure is quite uniform and very well aligned. This indicates a high quality of the epitaxial BiYIG film. The stripe domain period is around $480$~nm ($\pm 20$~nm) at remanence. The stripe-domain period decreases with increasing in-plane field, as shown in Fig.~\ref{fig:StaticMagnetic}(b) (See also Supplementary Material~II). This reflects the continuous evolution from a well-developed up/down stripe pattern at low field to a weakly modulated, nearly in-plane state close to saturation. A similar reduction of the stripe period close to in-plane saturation has been reported in weak-stripe Permalloy films~\cite{Liu2022}. As the field increases, the magnetization acquires a larger in-plane component and the field favours the canted regions, including the wall centers, whose magnetization is aligned with the field. The distinction between domains and walls therefore becomes progressively less sharp: the out-of-plane modulation weakens, the walls broaden, and the characteristic modulation length decreases. This also explains the weaker MFM contrast at fields near saturation. 

\subsection{Magnetization dynamics}

\begin{figure*}
	\centering
        \includegraphics[width=17cm]{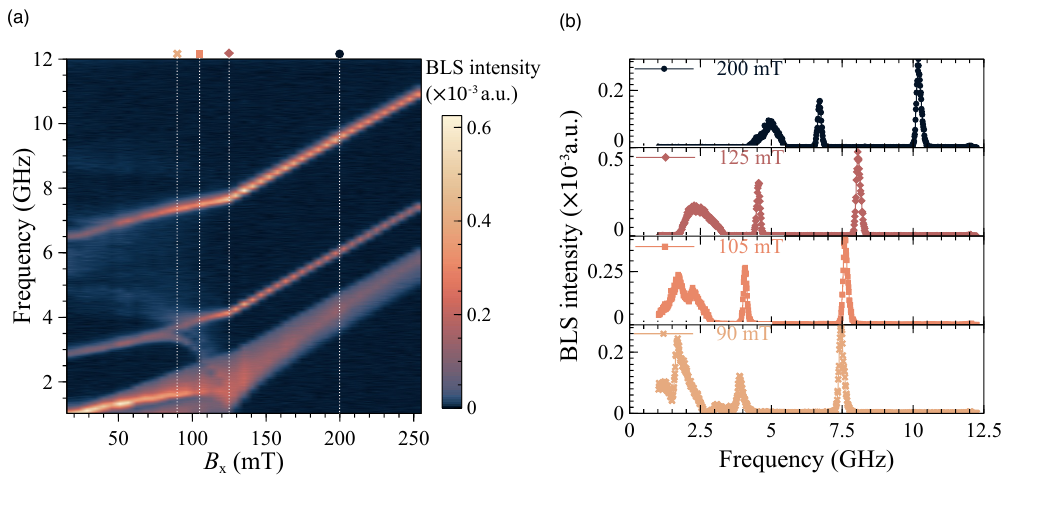}
	\caption{Spin-wave spectrum evolution. (a) Thermal $\mu$-BLS spectra measured over a frequency range of 1~GHz to 12~GHz with the applied in-plane field swept from 240~mT to 14~mT.(d) Line cuts for fixed in-plane field values: 200~mT, 125~mT, 105~mT and 90~mT. Three modes are observed above saturation . We observe mode softening at around 110~mT, with a mode that goes below the minimum measurable frequency and new mode emerging at the same field . This mode is seen also seen in the line cut at 105~mT. Many new modes emerge in the stripe domain phase.}
	\label{fig:exp-dynamics}
\end{figure*}

In order to map the reciprocal space evolution of the magnetization or in other words the spin wave spectra across uniform to stripe domain transition, we measure the thermal mode population using $\mu$-BLS on our BiYIG sample. Here, dc in-plane field is swept from positive field values when the sample is saturated down to very low fields where the stripe domain is stable, and the BLS intensity is recorded in the frequency range of 1~GHz to 12~GHz. The optical numerical aperture of $\mu$-BLS sets the detection limit of the k-vector to be around 20 rad/µm. The frequency versus field map of the BLS intensity is shown in Fig.~\ref{fig:exp-dynamics}(a), and line cuts at fixed fields are shown in Fig.~\ref{fig:exp-dynamics}(b)

In the saturated regime, we observe three main branches. Two of them are relatively narrow, while the lowest-frequency branch is clearly broader. The narrow higher-frequency branches suggest only weak frequency variation over the k-range probed by $\mu$-BLS, which is consistent with perpendicular standing spin-wave (PSSW)-like modes, also known as thickness-quantized modes. By contrast, the lowest branch is much broader, indicating a stronger dispersion over the wavevector window sampled by $\mu$-BLS. This low-frequency branch is therefore associated with the fundamental dipole-exchange mode, with a mixed Damon--Eshbach-like, backward-volume-like, and quasi-uniform character depending on the propagation direction. 

Now, as we decrease the in-plane field toward the critical fields, the lowest branch broadens further and shifts down towards the lowest experimentally accessible frequency. This is the signature of mode softening. If we look closely at the transition field (see Fig.~\ref{fig:exp-dynamics}(b) at $105$~mT), a new mode appears. This mode then increases in frequency with decreasing field while the first mode is seen to be pinned, with a finite amplitude at the lowest accessible frequency range (see Fig.~\ref{fig:exp-dynamics}(b)). On further reducing the field, we see that the spectrum gets richer, with continuing branches of the upper PSSW modes accompanied by new modes with broad amplitudes. 

The emergence of a low-frequency branch pinned at the experimental cutoff, together with another branch that rises in frequency, in the stripe phase, is suggestive of the soft-mode splitting expected for a periodically modulated magnetic state. This phenomenology resembles the Goldstone- and Higgs-like excitations reported in the stripe-domain transition in metallic systems\cite{grassi22}, although a definitive assignment would require direct mode-profile analysis.

\subsection{Dispersion relations and $\mu$-BLS model}

\begin{figure} 
	\centering
        \includegraphics[width=8.5cm]{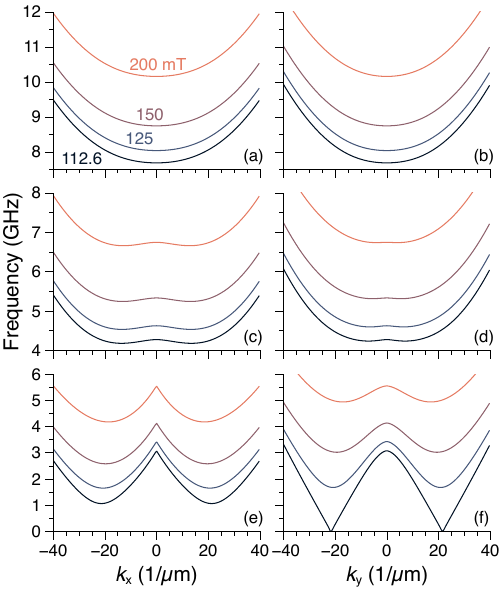}
	 \caption{Spin wave dispersion relations in the saturated state for different applied fields, $B_x$. Propagation along the field direction $x$ (a,c,e) and perpendicular to this direction along $y$ (b,d,f) are shown for the third-order (a,b), second-order (c,d) and first-order (e,f) PSSW modes.}
	\label{fig:disprel}
\end{figure}

In order to interpret the thermal $\mu$-BLS spectra, we calculate the dispersion relations at selected magnetic fields in the saturated state, using the magnetic parameters optimized using micromagnetic simulations (More details are provided in the next section). We use the dynamical matrix approach~\cite{Noertemann93, Henry16, Koerber21}, as implemented in the \textsc{TetraX} code~\cite{TetraX}, which involves solving the eigenvalues and eigenvectors of the linearized Landau-Lifshitz equation. Dispersion relations for wave propagation parallel to the DC field ($\mathbf{k}\parallel\mathbf{M}$) and perpendicular to it ($\mathbf{k}\perp\mathbf{M}$) are shown in Fig.~\ref{fig:disprel} for several values of the applied field.
The first- and second-order PSSW modes, as shown in Figs.~\ref{fig:disprel}(a)-\ref{fig:disprel}(d), exhibit a relatively flat dispersion within the experimental zone of interest ($|k| \lesssim 20$~$\mu$m), \textit{i.e.} vanishing group velocities, which explains the narrowness of these peaks in the $\mu$-BLS spectra. The lowest branch, shown in Figs.~\ref{fig:disprel}(e),\ref{fig:disprel}(f), on the other hand, exhibits a strong dispersion indicating large group velocity and hence the broadness of the $\mu$-BLS peak as it captures a large wavevector span. Interestingly, the lowest branch in both propagation directions exhibits a strong backward character which means a negative group velocity, such that the minimum frequency is at a finite wavevector. This type of behavior has recently been theoretically predicted, showing a finite-$k$ precursor prior to stripe domain formation. It has been further elaborated that the PMA tends to render the dispersion relations in all directions somewhat isotropic~\cite{Jan23,Lesniewski26} especially near fields close to transition, because the PMA-induced torque partially compensates the usual Zeeman, exchange, and dipolar restoring torques. As the field is reduced, the lowest branch of the dispersion relation with the wavevector perpendicular to the magnetization and hence the dc field, \textit{i.e.} the DE mode, softens at a finite critical wavevector, which in our case is around $k_c \sim 21$~rad/$\mu$m. This wavevector gives precise information on the periodicity of the stripe domains that form at the transition, which matches reasonably well the experimentally measured stripe domain periodicity close to the saturation field experimentally observed to be around $19.6$~rad/$\mu$m. The slight discrepancy is due to the fact that it is very difficult to observe the stripe domains and their exact periodicity at the transition using MFM since the majority of the moments are in plane. The slight out-of-plane modulation, a skeleton of the stripe domains does not provide enough signal over noise.

In order to interpret the measured $\mu$-BLS intensities and spectral line shapes of the modes observed in the saturated state, we used the formalism developed in the Ref.~\cite{Benaziz2025}. This approach sums up the back-scattering events collected within the numerical aperture of the microscope objective and accounts for both the magnetic mode profiles and the optical properties of the sample.

\begin{figure} 
    \centering
    \includegraphics[width=8.5cm]{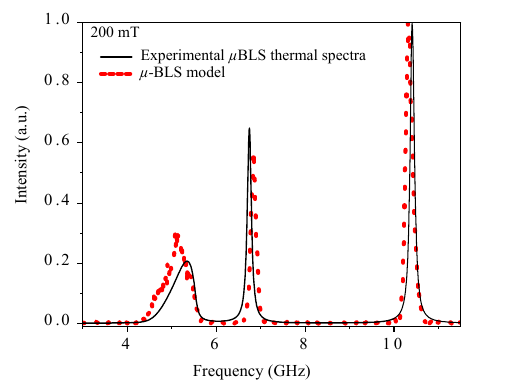}
    \caption{Experimental (red dots) and calculated (solid lines) $\mu$-BLS spectra at $\mu_0 H = 200~\mathrm{mT}$ for refractive index of $N=2.65+0.09i$}
    \label{fig:SignalVSmodel}
\end{figure}

As shown in Fig.~\ref{fig:SignalVSmodel}, the measured $\mu$-BLS intensity at $\mu_0 H = 200~\mathrm{mT}$ increases with the PSSW mode number. This trend is not expected from the magnetic mode profiles alone: the lowest mode, which corresponds to the nearly uniform Kittel mode at $k=0$ and to Damon--Eshbach or backward-volume spin waves at finite wavevector, would be expected to give the largest contribution because of its almost uniform thickness profile. However, the $\mu$-BLS intensity also depends on the optical properties of the sample, described by the complex refractive index $N$.

Since garnet films are transparent at the probe wavelength ($\lambda= 532$~nm), the dynamic magnetization throughout the full film thickness contributes coherently to the scattered amplitude. Light scattered from different depths therefore accumulates different optical phases, and the measured intensity is governed by a phase-weighted overlap between the spin-wave thickness profile and the optical field. This optical phase weighting can partially suppress the apparent intensity of the nearly uniform mode, while higher-order PSSW modes may exhibit a larger overlap integral and therefore a stronger $\mu$-BLS signal. Further details can be found in see Supplementary Material~III.

The spectrum calculated using this formalism is compared with the experimental data in Fig.~\ref{fig:SignalVSmodel} for $N=5.28+0.16i$. The model reproduces the mode positions, line shapes, and relative intensities of the PSSW modes reasonably well. The refractive index was treated as an effective fitting parameter, since literature values of $N$ lead to an underestimation of the intensity of the first peak, as discussed in the Supplementary Material~III.

The spectral line shapes are governed by the dispersive character of the corresponding spin-wave branches. The first and the second order PSSW modes exhibit nearly flat dispersions, giving rise to sharply peaked densities of states and therefore narrow spectral features. By contrast, the more dispersive Damon--Eshbach and backward-volume branches spread their spectral weight over a broader frequency range, leading to wider contributions in the $\mu$-BLS spectrum.

\subsection{Micromagnetic simulations}
We performed simulations with the finite-difference micromagnetic code \textsc{Mumax3}~\cite{vansteenkiste14} in order to gain better insight into the static and dynamic properties of our sample. The BiYIG film is modelled using a discretization scheme involving 512$\times$512$\times$8 finite difference cells, which nominally represents a system with physical dimensions of $L \times L \times 115$ nm. Periodic boundary conditions are applied within the film plane to mimic a continuous film and minimize finite-size effects. In the stripe phase, $L$ is varied such that the integer number of stripes within the simulation box corresponds to the equilibrium stripe domain period, while in the uniformly-magnetized state, $L$ is taken to be 8 $\mu$m. We assume a saturation magnetization of $M_s = 153$~kA/m, an exchange constant of $A = 4.4$~pJ/m, a perpendicular uniaxial anisotropy of $K_u = 15 259$~J/m$^3$, and a Gilbert damping constant of $\alpha = 10^{-3}$. We note that these parameters give an effective anisotropy field of $\mu_0 M_{\mathrm{eff}} = -7.2$ mT, which corresponds to the value determined from FMR experiments. The static magnetic field $H_x$ is applied along the $x$ direction, with the stripe domains running perpendicularly in the film plane along the $y$ direction.

\begin{figure} 
	\centering
        \includegraphics[width=8.5cm]{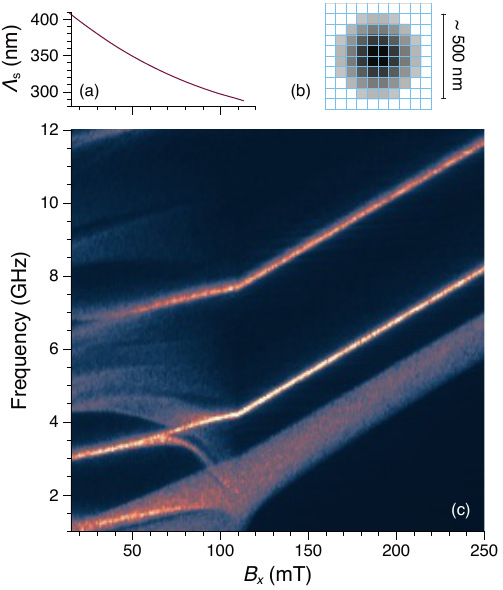}
	\caption{(a) Simulated equilibrium stripe domain period, $\Lambda_s$, as a function of applied field, $B_x = \mu_0 H_x$. (b) Schematic of Gaussian spot profile (gray levels) used for averaging output magnetization over supercells (light blue). (c) Color map of the simulated power spectral density of thermal spin waves as a function of $B_x$.}
	\label{fig:Simdyn}
\end{figure}

Below the saturation field, the stripe domain configuration is determined as follows. For each value of $H_0$, we initialize the system with $L$ = 8 $\mu$m and different integer numbers of stripe domains running along $y$, separated by Bloch walls magnetized along $x$, and let each configuration relax under the applied field. We then calculate the total magnetic energy of each relaxed configuration, which allows us to plot out the energy as a function of the integer number of periods. We then fit a quadratic function around the energy minimum to determine the true equilibrium stripe period, $\Lambda_s(H_0)$. Based on this value, we set the system size $L = N_s \Lambda_s(H_0)$ with $N_s$ being the integer value that brings $L$ as close as possible to 16~$\mu$m. For applied fields above the saturation field, where the ground state is the uniformly-magnetized, we use $L=16$~$\mu$m. The calculated stripe domain period as a function of applied field is shown in Fig.~\ref{fig:Simdyn}(a), which shows good qualitative agreement with the experimental observations.

In order to estimate the power spectrum of thermal spin waves in simulation, we include a random, thermal field in the effective field of the Landau-Lifshitz-Gilbert equation with a temperature of $T = 300$~K in order to populate all spin wave modes~\cite{Leliaert17}. We then construct the output magnetization as follows. We record the $z$-component of the magnetization (the component most strongly probed by BLS) averaged over a series of ``supercells''~\cite{Massouras24}, where each supercell $i$ represents $4 \times 4 \times 1$ finite-difference cells at the top surface of the film. Within the region of interest, as shown in Fig.~\ref{fig:Simdyn}(b), we sum $m_{z,i}$ over all supercells with a weighting factor given by a Gaussian function, with a full-width at half-maximum of 500 nm to mimic the experimental laser spot size. This then produces a time series in the output quantity, $\langle m_z(t) \rangle$. The power spectral density is then estimated using the Welch method~\cite{Welch67}. Specifically, we record $\langle m_z(t) \rangle$ over a duration of 200~ns, which is then subdivided into half-overlapping 20-ns Hann windows, resulting in a frequency resolution of 50~MHz that approximates the instrumentation limit of our experiment. The power spectrum is then computed for each window from a discrete Fourier transform of $\langle m_z(t) \rangle$, which is then averaged over all windows.

Color map of the simulated power spectral density of thermal spin waves as a function of in-plane field is shown in Fig.~\ref{fig:Simdyn}(c). We observe three main modes in the saturated state as we did in the experiments. We observe the softening of the lowest mode close to an in-plane field of $110$~mT which agrees well with the experiments. Furthermore, at the transition, we see new modes appearing as we enter the stripe domain state reproducing the experimental spectra remarkably well. This shows that our measured magnetic parameters and our interpretation of the dispersion relations and the $\mu$-BLS spectra are consistent. Overall, the simulations show excellent agreement with the experimental results. 

\section{Discussion}

Our results connect several branches of earlier work on stripe-domain-based magnonics. Previous studies on metallic multilayers have shown that stripe-domain textures can act as magnonic crystals, producing folded and hybridized spin-wave bands. More recent work on reconfigurable Pt/Co stripe-domain systems has further demonstrated that the propagating spin-wave spectrum strongly depends on the orientation of the stripes with respect to the propagation direction. The present work extends this picture by focusing on the dynamical pathway into such a texture-defined spectrum, following the softening and reorganization of thermal spin-wave modes across the saturation-to-stripe transition in a low-damping insulating BiYIG film.

A key observation is the quantitative agreement between the real-space stripe period and the calculated softening wave vector. Close to the transition, the stripe period measured by MFM is around $\Lambda_{s}\simeq 320~\mathrm{nm}$, corresponding to $k_{\rm stripe}=2\pi/\Lambda_{s}\simeq 19.6~\mathrm{rad}/\mu\mathrm{m}$.  This value agrees well with the calculated Damon--Eshbach softening minimum near $k_{\rm DE}\simeq 20$--$21~\mathrm{rad}/\mu\mathrm{m}$. Such agreement is consistent with the spin-wave-freezing scenario, in which the critical finite-$k$ excitation of spin wave selects the period of the stripe modulation below the transition. Since this softening occurs for $\mathbf{k}\perp\mathbf{M}$, the modulation develops perpendicular to the in-plane magnetization, and the stripe domains are aligned along this direction.

The observed stripe-state spectrum also connects to earlier reports of collective modes in weak-stripe domain systems. In particular, the low-frequency branch pinned near the experimental cutoff, and the branch that appears and rises in frequency sharply in the stripe phase, resembles the phase- and amplitude-like mode phenomenology respectively, reported earlier for metallic systems, which were assigned to Golstone and Higgs like modes. Compared with previous studies that mainly 
addressed either spin-wave bands in an already formed stripe texture or the theoretical 
freezing mechanism, our work experimentally connects the real-space stripe period, the 
finite-$k$ softening wave vector, and the thermal $\mu$-BLS spectral reorganization in a 
low-damping insulating PMA film.

\section{Conclusion}
In conclusion, we have investigated the spin-wave dynamics across the uniform-to-stripe transition in BiYIG films with perpendicular magnetic anisotropy using magnetic imaging, microfocused BLS spectroscopy, and micromagnetic simulations. The approach to the transition is marked by the softening of a finite-wave-vector mode in the saturated phase, while the stripe phase exhibits a dense and reorganized spectrum of collective excitations. Micromagnetic simulations reproduce the main experimental $\mu$-BLS spectra across the transition, and the calculated BLS intensities and mode frequencies at selected fixed fields show good agreement with the measured spectra. This agreement confirms that the observed spectral evolution reflects the intrinsic spin-wave modes of the field-dependent magnetic configuration.
Our results support a picture in which PMA strongly reshapes the low-frequency dispersion and reduces the usual distinction between Damon--Eshbach and backward-volume geometries close to the transition. Nevertheless, the finite-k Damon--Eshbach minimum provides the relevant instability: it softens as the in-plane field is reduced and freezes into the static stripe modulation below the critical field. The emergence of a low-frequency branch pinned near the experimental cutoff, together with a higher-frequency branch that hardens in the stripe phase, is suggestive of the soft-mode reorganization expected after translational symmetry breaking, although a definitive Goldstone/Higgs assignment requires direct mode-profile analysis. These findings establish BiYIG stripe domains as a low-damping, field-tunable platform for exploring spin-wave freezing, magnetic phase transitions, and self-assembled magnonic crystals.

\begin{acknowledgements}
This work was partially supported by the Agence Nationale de la Recherche (France) under contract numbers  ANR-22-CE30-0014 (DeMIuRGe) and ANR-24-CE24-0354 (ANTIPASTI). We express our sincere gratitude to Grégoire de Loubens and Nicolas Vukadinovic for fruitful discussions and, André Thiaville and Stanislas Rohart for their help during SQUID-VSM and MFM measurements. 
\end{acknowledgements}

\bibliography{BiYIG}

\clearpage
\onecolumngrid

\section*{Supplemental Material}
\setcounter{figure}{0}
\renewcommand{\thefigure}{S\arabic{figure}}

\setcounter{table}{0}
\renewcommand{\thetable}{S\arabic{table}}

\setcounter{equation}{0}
\renewcommand{\theequation}{S\arabic{equation}}

\setcounter{section}{0}
\renewcommand{\thesection}{S\Roman{section}}

\section{Magnetic Parameters from FMR Measurements}

\begin{figure}[h]
\centering
\includegraphics[width=\linewidth]{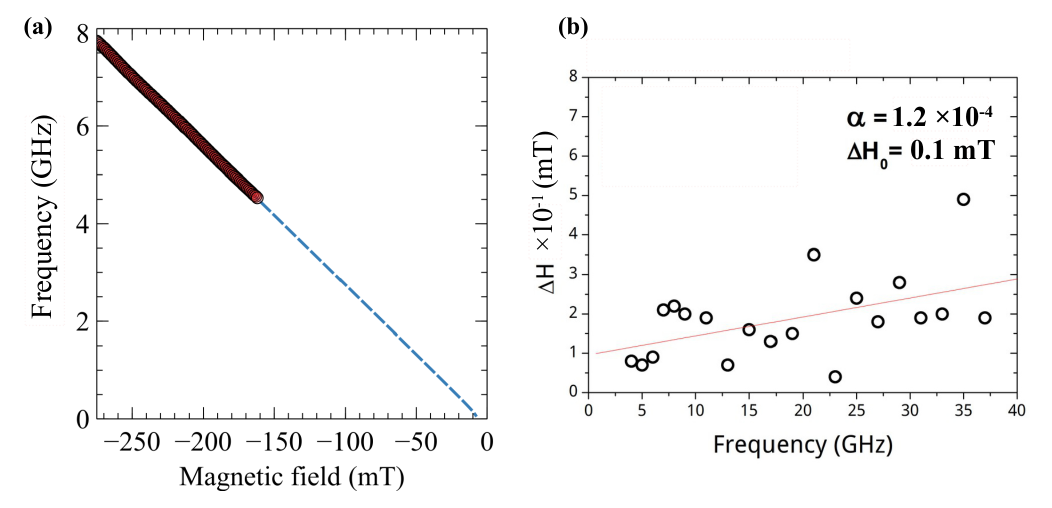}
\caption{(a) VNA-FMR measurement of the BiYIG sample with thickness $t = 120$ nm. The red dots represent the measured frequencies of the Kittel mode in the saturated state, while the blue dashed line shows the fit to the Kittel equation, yielding an effective magnetization of $-7.2 \pm 0.5$~mT. (b) Extracted linewidth as a function of the field.}
\label{fig:Meffanddamping}
\end{figure}

FMR resonance measurements were performed with a Coplanar Waveguide (CPW) connected to a vector network analyzer.  The sample was placed between two electromagnetic coils for the DC field to be in-plane, the DC field was swept from $-275$ to $275$~mT by step for $1$~mT, the RF current was swept from $4$~MHz to $16$~GHz with steps of $4$~MHz, reflection and transmission coefficients were acquired. The input power of the microwaves field was set to $-20$~dBm to avoid any non-linear processes.
The magnetic susceptibility coefficients were constructed from the raw data measurement~\cite{Bilzer2007}.
The resonant frequencies of the Kittel mode in the saturated states were then identified. The resonant Kittel frequencies as a function of the in-plane applied field were then fitted to find the effective magnetization as shown in Fig~\ref{fig:Meffanddamping}(a).

The fitting function is the classical Kittel formula for thin film :
\begin{equation}
    f^2=(\mu_0\gamma/2\pi)^2H_0(H_0+M_ \mathrm{{eff}}),
\end{equation}
with $\mu_0$ the free space magnetic permeability in T.m/A, $\gamma/2\pi$ the gyromagnetic ratio in Hz/T, $H_0$ the applied magnetic field in A/m and $M_ \mathrm{{eff}}$ the effective magnetization in A/m.

The damping $\alpha=1.2\times10^{-4} $is extracted using 
\begin{equation}
 \mu_0 \Delta H_{\mathrm{FWHM}}
=
\mu_0 \Delta H_0
+
\frac{4\pi\alpha}{\gamma}\,f
\end{equation}

Where $\mu_0 \Delta H_0 = 0.1$~mT is the inhomogeneous broadening and $\Delta H_{\mathrm{FWHM}}$ is the full width at half maximum of the resonance peak as shown in Fig~\ref{fig:Meffanddamping}(b).

\section{Magnetic force microscopy and stripe periodicity}

\begin{figure}[h]
    \centering
    \includegraphics[width=\linewidth]{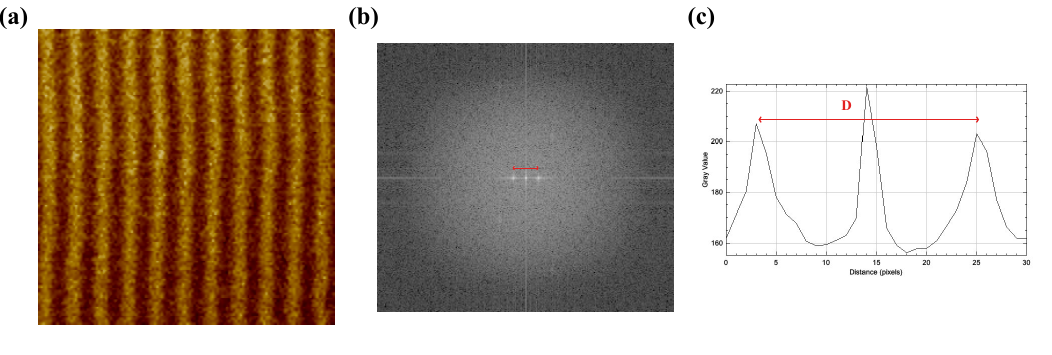}
    \caption{Real space MFM image of the stripe at remanence and its equivalent spectral density image in the Fourier space.}
    \label{fig:MFMcalc}
\end{figure}

Magnetic force microscopy (MFM) images were acquired using a Park Systems atomic force microscope equipped with a Caylar magnetic-field module and cobalt-coated tips. Images of $5\times5~\mu\mathrm{m}^{2}$ were recorded with $256\times256$ pixels at a scan rate of $1~\mathrm{Hz}$. The lift height was varied between $70$ and $160~\mathrm{nm}$ to verify that the magnetic configuration was not perturbed by the tip, and a fixed lift height of $100~\mathrm{nm}$ was used for the field-dependent measurements.

The average stripe-domain periodicity was extracted from the two-dimensional fast Fourier transform of the MFM images using ImageJ as shown in Fig.~\ref{fig:MFMcalc}. The stripe modulation produces two satellite peaks at $\pm k_0$ around the central peak, where $k_0$ is expressed in $\mathrm{rad/\mu m}$. The stripe period is given by

\begin{equation}
    P=\frac{2\pi}{k_0}
    =\frac{4\pi}{\Delta k},
\end{equation}

where $\Delta k=2k_0$ is the separation between the two satellite peaks. For an image of lateral size $L$, the reciprocal-space sampling interval is $\delta k=2\pi/L$. Equivalently, if the satellite peaks are separated by $D$ FFT pixels, the period is $P=2L/D$.
A representative extraction is shown in Fig.~\ref{fig:MFMcalc}: the two satellite peaks are separated by approximately $D=21$ FFT pixels, corresponding to $\Delta k=D(2\pi/L)\simeq26.4~\mathrm{rad/\mu m}$ and hence to $k_0=\Delta k/2\simeq13.2~\mathrm{rad/\mu m}$, yielding a stripe periodicity $P=2\pi/k_0\simeq0.48~\mu\mathrm{m}= 480$~nm.

\section{Amplitude interpretation of the $\mu$-BLS signals}

\begin{figure}[h]
    \centering
    \includegraphics[width=\linewidth]{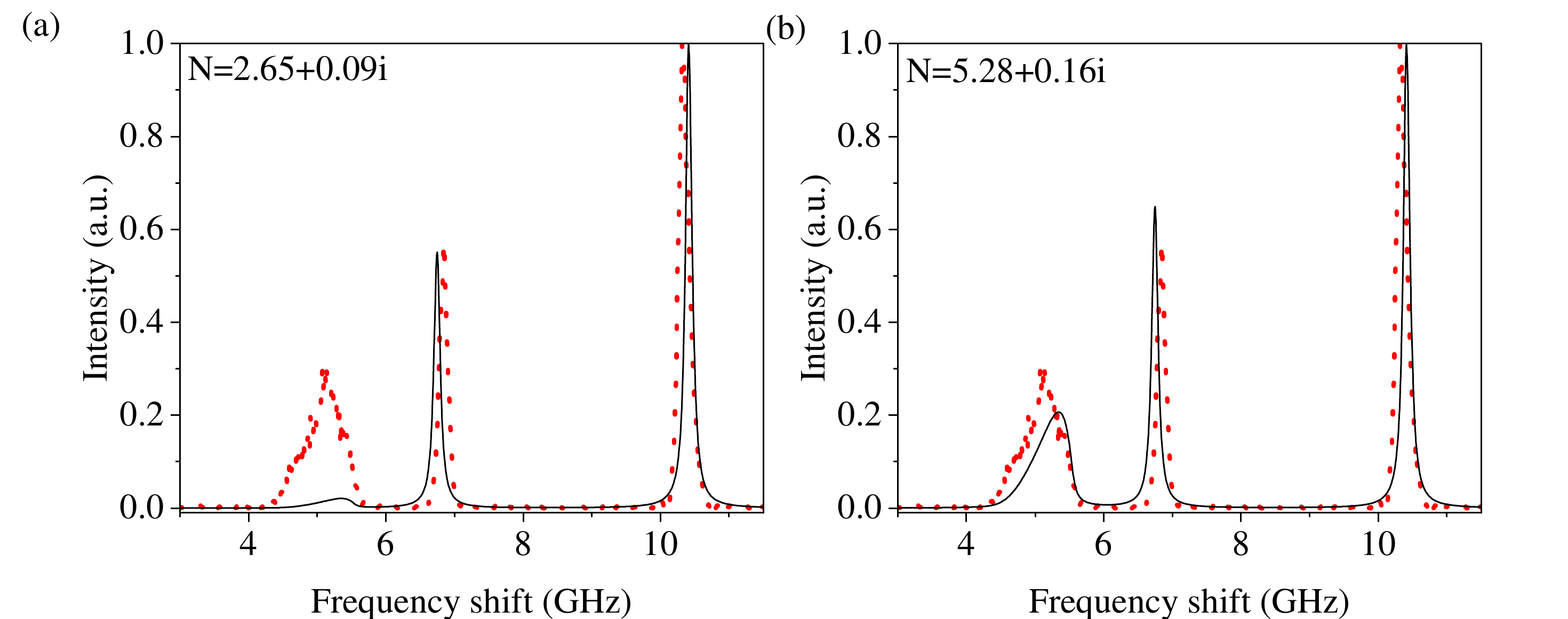}
    \caption{Comparison of the experimental (solid black) and calculated (red dotted) $\mu$-BLS spectra at 200 mT for two different refractive indices (a) N= $2.65+0.09i$ and (b) $5.28+0.16i$  }
    \label{fig:BLSmodelcomp}
\end{figure}

\begin{figure}[h]
    \centering
    \includegraphics[width=\linewidth]{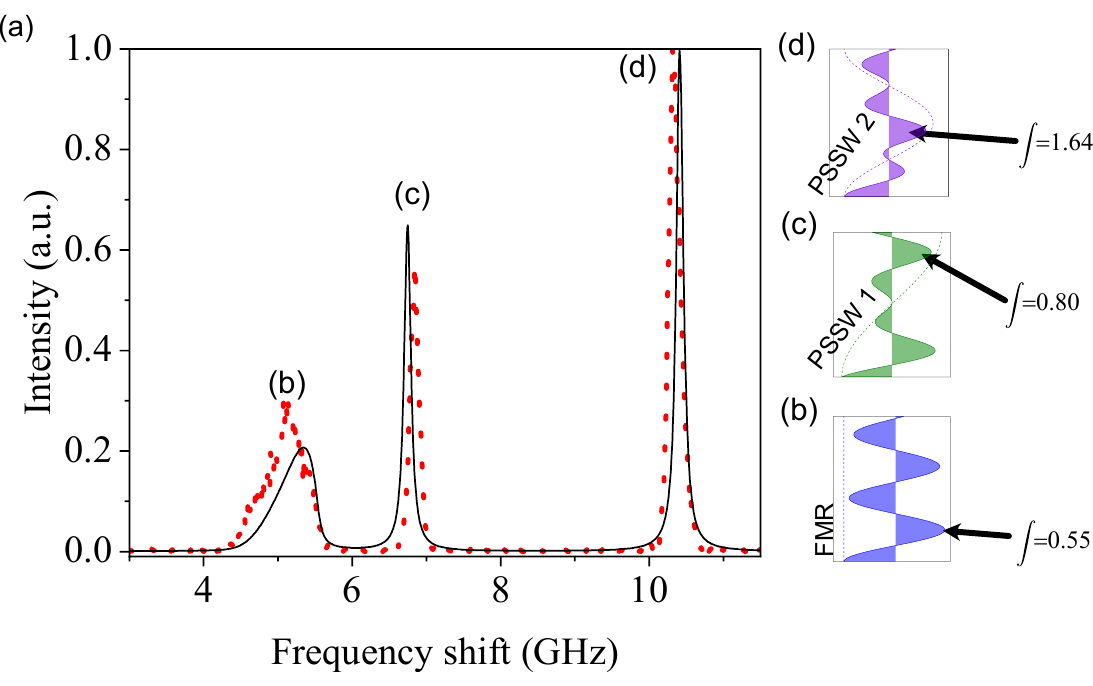}
     \caption{Experimental and calulated $mu$-BLS spectra at 200 mT showing three modes and there respective Thickness profiles at k=0 in dotted lines and the overlap functions with N= $5.28+0.16i$ showing the resulting integral.}
    \label{fig:magprofandopticalphase}
\end{figure}

The calculations were performed using the micromagnetic parameters, using a refractive index of $N = 2.65+0.09i$, and an off-diagonal permittivity of $0.029+0.01i$ taken from the literature~\cite{Jesenska16}.
Overall, the calculated spectrum closely reproduces the experimental spectral positions, line shapes, and relative intensities, confirming the accuracy of the extracted micromagnetic parameters. However, the model underestimates the amplitude of the first peak as shown in Fig.~\ref{fig:BLSmodelcomp}(a). This discrepancy could originate from the fact that the optical formalism is derived under the ultrathin approximation, which assumes the film thickness to be much smaller than the optical penetration depth $\delta = \lambda/(4\pi\,\mathrm{Im}(n))$ (here $\delta \sim 265$ nm), and neglects multiple internal reflections and the resulting interference of the optical field within the film.
For the 115 nm Bi-YIG film, this condition is not fulfilled. As a result, the first peak, which is nearly uniform across the thickness, is particularly sensitive. To account for this effect, we introduce an effective refractive index $n_\mathrm{eff}$ as a phenomenological fitting parameter. It should not be interpreted as a physical material constant, but rather as a modification of the optical phase accumulation within the model. In Fig.~\ref{fig:BLSmodelcomp}(b), the model calculated using $N = 5.28+0.16i$ improves the agreement with the experimental amplitude without modifying the peak frequencies and line shapes, as the spectral line shapes are governed by the dispersive characteristics of the spin wave modes.

Furthermore, the origin of the peak shape comes from the magnetic properties of the sample, while the origin of the increasing mode intensities with mode number comes from the interplay between the mode profiles across the thickness with the electromagnetic wave. In a simplified description, the BLS amplitude can be written as
\begin{equation}
    A_\mathrm{\mu BLS}=\int dz~m_z(z) \exp(iz\frac{4\pi N}{\lambda})
    \label{amplitudeBLS}
\end{equation}
where N is the complex refractive index and $\lambda$ is the optical wavelength.
The sample is transparent at the wavelength used for the BLS measurements. The real part of N accounts for the phase accumulation, while its imaginary part accounts for optical absorption.


Therefore, incident and scattered electromagnetic waves probe the entire thickness of the film. Fig.~\ref{fig:magprofandopticalphase}(b,c,d) shows the thickness profiles of the FMR mode and of the first two PSSW modes in dotted lines at k=0. Based only on the magnetic profiles, one would expect the uniform FMR mode to exhibit the strongest BLS intensity, since its dynamic magnetization has the same sign throughout the whole thickness. In contrast, the PSSW modes exhibit nodes across the thickness, leading to partial cancellation of the magnetic contribution when integrated over the film thickness.
However, the calculated intensities shown in Fig.~\ref{fig:magprofandopticalphase}(a) reveal the opposite trend. This behavior originates from the optical phase factor. Fig.~\ref{fig:magprofandopticalphase}(b,c,d) in blue green and purple respectively, shows the results of  Eq. \ref{amplitudeBLS} using the effective refractive index, which takes into account the magnetization profile as well as the optical phase factor into account. For the uniform mode, the magnetic contribution is identical throughout the thickness, but the light scattered from different depths accumulates different optical phases. The resulting partial destructive interference reduces the BLS intensity and gives a low integral. In contrast, the spatial variation of the PSSW modes can partially compensate for this optical phase rotation, leading to a larger overlap integral and therefore to a stronger BLS signal. The values of the overlap integrals (calculated by taking into account both the real and imaginary part of the magnetization profile and the full k-window of $\mu$-BLS) confirm this interpretation. The second PSSW mode yields a larger overlap integral than the FMR mode, resulting in a stronger BLS intensity.

\end{document}